\documentclass[11pt]{article}

\usepackage[preprint]{acl}

\usepackage{times}
\usepackage{latexsym}

\usepackage[T1]{fontenc}

\usepackage[utf8]{inputenc}

\usepackage{microtype}

\usepackage{inconsolata}

\usepackage{graphicx}
\usepackage{float}
\usepackage{stfloats}
\usepackage{booktabs}
\usepackage{tabularx}
\usepackage{multirow}
\usepackage{url}
\usepackage{xurl}

\title{DocPC: Document-Level Visual Retrieval via Representative Page Composition}

\author{
  \textbf{Chengsong You\textsuperscript{1,3}},
  \textbf{Qiyi Jiang\textsuperscript{1}},
  \textbf{Junwei Zhou\textsuperscript{2}\thanks{Corresponding authors.}},
  \textbf{Xiaoyu Cao\textsuperscript{1}},
  \textbf{Weiyao Wang\textsuperscript{1}},
\\
  \textbf{Yiwei Xu\textsuperscript{1,3}},
  \textbf{Ziyan Zhao\textsuperscript{1}},
  \textbf{Zhen Sun\textsuperscript{1,3}},
  \textbf{Qicheng Zhu\textsuperscript{4}},
\\
  \textbf{Xuanyi Fu\textsuperscript{5}},
  \textbf{Yufan Chen\textsuperscript{5}},
  \textbf{Yilun Li\textsuperscript{5}},
  \textbf{Rongkang Xiong\textsuperscript{5}},
\\
  \textbf{Yunhai Hu\textsuperscript{6}},
  \textbf{Nan Du\textsuperscript{2}\footnotemark[1]}
\\
\\
  \textsuperscript{1}Thin Red Line,
  \textsuperscript{2}Matter Innovation Inc.,
  \textsuperscript{3}East China Normal University,
\\
  \textsuperscript{4}UC Berkeley,
  \textsuperscript{5}Independent Researcher,
  \textsuperscript{6}New York University
\\
  \small{\texttt{51275901122@stu.ecnu.edu.cn, zjw330501@gmail.com, frankdu@matter.ai}}
}

\begin{document}
\maketitle
\begin{abstract}
Visual document retrieval has advanced by encoding page screenshots with vision-language models, bypassing OCR pipelines. However, existing methods remain page-centric, misaligned with real-world scenarios requiring complete document retrieval. A naive page-then-document aggregation suffers from linear indexing cost and degraded retrieval when relevance spans multiple pages. We propose \textbf{DocPC}, a document-level visual retrieval framework based on \textbf{Representative Page Composition}: selecting representative pages and composing them into a single grid image for document-level indexing, reducing indexed images, vectors, and storage by \textbf{10.1$\times$} and end-to-end indexing time by \textbf{roughly 7.7$\times$}. To handle multi-positive supervision prevalent at the document level, we combine \textbf{multi-positive contrastive learning} with \textbf{sparsely scheduled listwise optimization}. We also introduce \textbf{DocViRe}, a benchmark with multi-positive relevance annotations. DocPC-ColQwen achieves NDCG@5 of \textbf{44.09} on DocViRe, outperforming the strongest page-level baseline at \textbf{38.91} while reducing storage by \textbf{10.1$\times$}. Code is available at \url{https://anonymous.4open.science/r/DocPC-Document-Level-Visual-Retrieval-via-Representative-Page-Composition-1D52}. Data is available at \url{https://huggingface.co/datasets/anonymous-7219/docpc}.
\end{abstract}

\section{Introduction}

Visual document retrieval has emerged as a compelling alternative to OCR-based pipelines by directly encoding document page screenshots with vision-language models \citep{faysse2024colpali}. Models such as ColPali and ColQwen represent page images as multi-vector embeddings and perform late-interaction matching \citep{khattab2020colbert,santhanam-etal-2022-colbertv2}, achieving strong results on page-level benchmarks such as ViDoRe.

However, existing visual document retrieval remains largely \textbf{page-level}. While page-level retrieval suits retrieval-augmented generation \citep{lewis2020retrieval}, many practical scenarios, such as enterprise search, academic retrieval, and legal review, require retrieving \textbf{complete documents}. Yet cross-document \textbf{document-level} visual retrieval remains underexplored. ViDoRe \citep{faysse2024colpali,mace2025vidore2} and MMDocIR \citep{dong2025mmdocir} focus on page-level retrieval, while VisDoMBench \citep{suri2025visdombench} targets QA rather than retrieval.

A natural fallback is indexing pages independently and aggregating scores to the document level (e.g., MaxP). This faces a \textbf{dual challenge}: index size scales linearly with total pages, and relevance distributed across multiple pages may not be captured by a single-page score.

We propose \textbf{DocPC}, a document-level visual retrieval framework built upon page-level retrievers via supervised fine-tuning. Its core idea, \textbf{Representative Page Composition (PC)}, selects $K$ representative pages and arranges them into a single grid image, reducing document-side indexing from $\mathcal{O}(N)$ to $\mathcal{O}(1)$ without additional model complexity.

Document-level retrieval also amplifies the \textbf{training challenge} of multi-positive supervision: topically related documents are more common, making standard InfoNCE \citep{oord2018infonce} prone to treating co-relevant documents as negatives. We combine \textbf{Multi-Positive Contrastive Loss} \citep{khosla2020supervised,rivas2024conked} with a sparsely scheduled ApproxNDCG-based listwise loss \citep{qin2010general} to address this.

We further construct \textbf{DocViRe}, a benchmark for document-level visual retrieval with multi-positive relevance annotations across seven domains. Our contributions:
\begin{itemize}
    \item We formalize document-level visual retrieval and propose Representative Page Composition for $\mathcal{O}(1)$ document indexing.
    \item We construct DocViRe, a dedicated document-level visual retrieval  benchmark with multi-positive annotations.
    \item We empirically validate DocPC on DocViRe, demonstrating consistent retrieval gains from page selection and multi-positive training while significantly reducing index size.
\end{itemize}
\section{Related Work}

\subsection{Visual Document Retrieval}

Visual document retrieval has advanced rapidly by encoding document page images with vision-language models (VLMs) and matching queries via late interaction over token-level embeddings \citep{faysse2024colpali,khattab2020colbert,santhanam-etal-2022-colbertv2}. ColPali is a representative OCR-free framework in this line: it produces multi-vector representations from page screenshots and applies late-interaction scoring, avoiding complex text extraction and ingestion pipelines while achieving strong retrieval performance on visually rich documents \citep{faysse2024colpali}. Subsequent work has strengthened the visual backbone, for example through ColQwen \citep{faysse2024colpali}, explored more efficient retrieval architectures such as DSE \citep{ma2024dse}, and integrated visual retrievers into end-to-end visual RAG systems such as VisRAG \citep{yu2025visrag}. To reduce per-page indexing cost, efficient variants such as Light-ColPali \citep{ma2025lightcolpali} and DocPruner \citep{yan2025docpruner} further reduce the number of document-side vectors. However, these methods still operate at the \emph{page} level: they index and rank individual pages rather than complete documents. In contrast, this work targets \emph{document-level} visual retrieval and represents an entire multi-page document with a single composed encoding.

\subsection{Retrieval Granularity and Benchmarks}

Existing benchmarks for visual document retrieval are largely page-centric. ViDoRe \citep{faysse2024colpali,mace2025vidore2} evaluates cross-document retrieval where the retrieval unit is a single page. MMDocIR \citep{dong2025mmdocir} focuses on intra-document page-level and layout-level retrieval. VisDoMBench \citep{suri2025visdombench} studies multi-document question answering rather than document ranking. As a result, current benchmarks do not provide a standard setting for \emph{document-level} visual retrieval, where the system must rank whole documents from a corpus of multi-page documents.

A natural alternative is to reuse ideas from text retrieval by indexing smaller units and aggregating them to the document level. Passage-to-document aggregation has been widely studied in text retrieval, where methods such as MaxP aggregate passage scores into document scores \citep{dai2019deeper,li2023parade}. Multi-page document understanding has also been extensively explored for long-document question answering and reasoning, for example in Hi-VT5 \citep{tito2023hivt5}, DUDE \citep{Van_Landeghem_2023_ICCV}, M-LongDoc \citep{chia-etal-2025-longdoc}, and DocOwl~2 \citep{hu2025docowl2}. Nevertheless, these settings differ from document retrieval: the goal is typically to answer questions over a given document rather than to rank documents by relevance.

To clarify the landscape, we distinguish three related but distinct task settings. (1) \emph{Cross-document retrieval} ranks whole documents from a corpus given a query; this is the setting we address. (2) \emph{Intra-document page retrieval} finds relevant pages within documents to support downstream tasks such as QA. M3DocRAG \citep{cho2024m3docrag} retrieves pages via ColPali embeddings and feeds them to a multimodal LM for question answering. The key difference from our setting is that these methods retrieve \emph{pages} rather than \emph{documents}, and retrieval serves as an intermediate step rather than the end task. (3) \emph{Multi-page document QA} evaluates comprehension over a given long document without corpus-level retrieval. MMLongBench-Doc \citep{ma2024mmlongbench} benchmarks LVLMs on long PDF documents with cross-page questions, and LongDocURL \citep{deng2025longdocurl} integrates understanding, reasoning, and locating across 33,000+ pages. Our work targets setting~(1), which remains underexplored compared to~(2) and~(3). To the best of our knowledge, there is still no benchmark or systematic study for cross-document \emph{document-level} visual retrieval with multi-positive relevance annotations.
\section{Task Formulation}
\label{sec:task_formulation}

\paragraph{Setting.} We consider \textbf{document-level} visual retrieval. The input is a text query $q$ and a corpus $\mathcal{D} = \{d_1, d_2, \ldots, d_M\}$ of $M$ documents. Each document $d_i$ is a multi-page visual object: $d_i = \{p_1^{(i)}, p_2^{(i)}, \ldots, p_{N_i}^{(i)}\}$, where $p_j^{(i)}$ denotes the $j$-th page of document $i$ (e.g., a page image). The system assigns a relevance score $s(q, d_i) \in \mathbf{R}$ to each document and returns a ranking over $\mathcal{D}$. The goal is to rank relevant documents above irrelevant ones, where relevance is defined at the \emph{document} level.

\paragraph{Differences from page-level retrieval.} Two issues distinguish document-level from page-level retrieval. (1) \textbf{Representation compression:} A document has a variable number of pages $N_i$. Efficient retrieval therefore requires a representation whose size does not grow linearly with $N_i$. Page-level methods index each page separately and incur $\mathcal{O}(\sum_i N_i \cdot T)$ storage (with $T$ vectors per page), which is prohibitive at scale. (2) \textbf{Multi-positive document supervision:} At the document level, many queries naturally have multiple relevant documents (e.g., several reports on the same policy). Training with a single positive per query mislabels other relevant documents as negatives and degrades learning. We therefore need objectives that support multiple positives per query.

\paragraph{Page-aggregation baseline.} A straightforward baseline aggregates page scores into a document score: $s(q, d_i) = \mathrm{Agg}_{p \in d_i}\, s(q, p)$ (MaxP). This does not reduce index size. When relevance is \emph{distributed} across pages, no single page has a dominant score, so MaxP tends to underrank such documents. DocPC instead builds one composite representation per document and scores it directly.

\begin{figure}[t]
  \centering
  \includegraphics[width=\columnwidth]{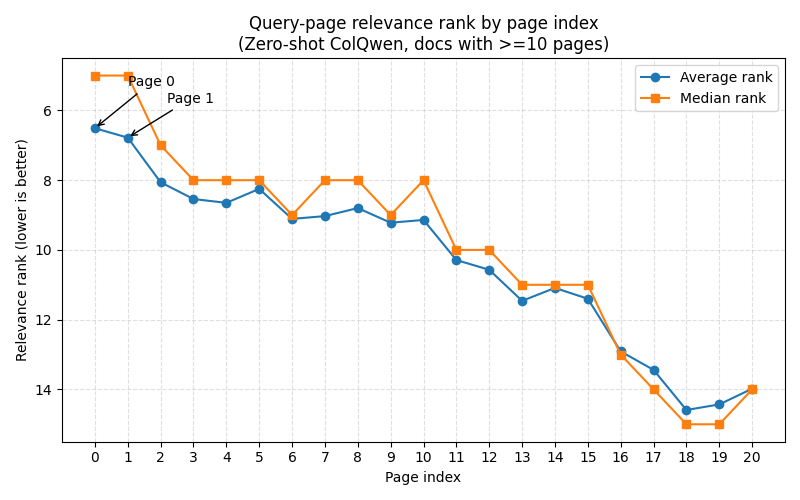}
  \caption{Average relevance rank by page index on a randomly sampled 10\% subset of the DocViRe training split. For relevant query-document pairs, we compute the page-wise relevance rank produced by zero-shot ColQwen and report the first 20 page positions. Lower rank indicates stronger relevance. The first two pages are ranked most favorably, suggesting that early pages contain strong document-level signals.}
  \label{fig:relevance-distribution}
\end{figure}

\paragraph{Relevance distribution analysis.} We analyze page-wise relevance ranks produced by zero-shot ColQwen on a randomly sampled 10\% subset of the DocViRe training split (Figure~\ref{fig:relevance-distribution}). The first two pages achieve the best average ranks, indicating that early pages often carry stronger document-level signals. This motivates early-page-aware selection strategies in DocPC.

\section{DocViRe Benchmark}
\label{sec:dataset}

Existing visual document retrieval benchmarks such as ViDoRe are defined at the \emph{page} level: queries typically target local evidence within a single page, and systems are evaluated by whether they retrieve the correct page. In contrast, the task studied here takes the whole document as the retrieval unit, where relevance depends on document-level topic, cross-page information, and global semantics. Page-level annotations in ViDoRe therefore cannot be directly reused as document-level ground truth. Moreover, existing benchmarks are not designed for the cross-document document-ranking setting considered in this work. We thus construct \textbf{DocViRe}, a benchmark explicitly designed for \emph{document-level} visual retrieval.

DocViRe is designed to support document-level retrieval with document-grounded queries and multi-positive relevance annotations, while covering cross-document retrieval across multiple domains and providing a reference training set for fine-tuning trainable retrievers.

Documents are collected from the public PDFA dataset \citep{pdfa_hf_2024}. We consider seven domains: biology, education, finance, government, industrial, legal, and research.

\begin{figure}[H]
  \centering
  \includegraphics[width=\columnwidth]{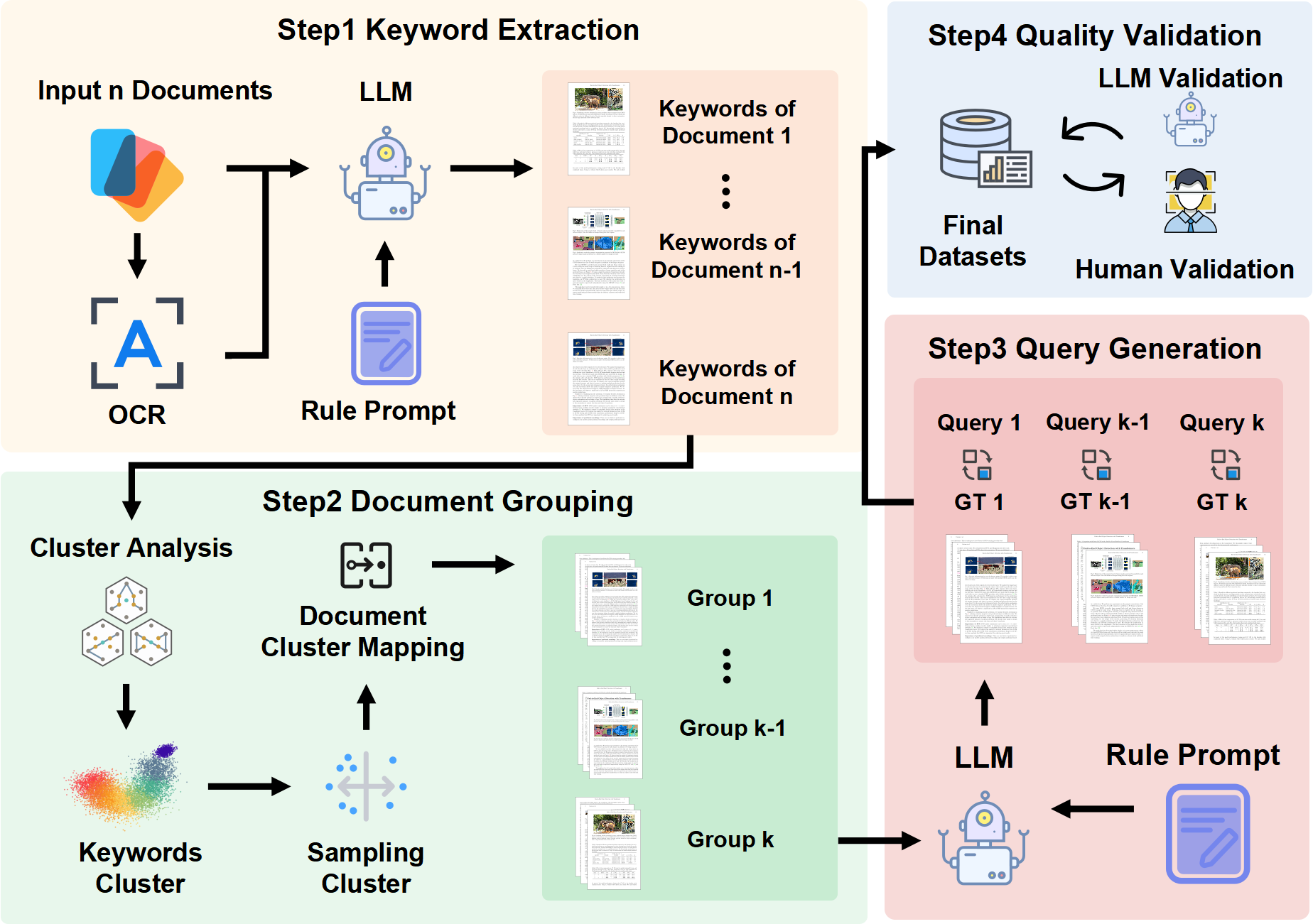}
  \caption{Pipeline for constructing DocViRe. Documents are first processed with OCR and a large language model (LLM) to extract keywords. Cluster analysis then groups documents by topical similarity and assigns each to a group. For each group, an LLM generates document-level query-ground-truth pairs, which are validated by a second LLM pass and by human review to produce the final benchmark.}
  \label{fig:dataset-pipeline}
\end{figure}

\begin{sloppypar}
Table~\ref{tab:dataset} reports the train and test split per domain, including query and document counts. In total, the training set contains 10,884 queries and 4,438 documents. The test set contains 2,237 queries and 3,850 documents. The training and test splits are constructed from two disjoint document pools with no document overlap, ensuring that evaluation is performed on unseen documents rather than re-queried instances from the training corpus. These splits are designed to support both fine-tuning of trainable retrievers and standardized evaluation for document-level retrieval. Queries and ground-truth document sets are obtained through the construction pipeline described below. Figure~\ref{fig:dataset-pipeline} summarizes that pipeline. We describe the four steps in the following paragraphs.
\end{sloppypar}
\medskip
\begin{table*}[t]
\centering
\footnotesize
\caption{DocViRe benchmark statistics per domain (training set and test set). The train and test splits are document-disjoint. All in English.}
\label{tab:dataset}
\renewcommand{\arraystretch}{1.05}
\setlength{\tabcolsep}{0.8em}
\begin{tabular*}{\linewidth}{@{}l@{\extracolsep{\fill}}r@{\hspace{1.5em}}r@{\hspace{2.5em}}r@{\hspace{1.5em}}r@{\hspace{2em}}>{\raggedright\arraybackslash}p{0.45\linewidth}@{}}
\toprule
\multirow{2}{*}{\textbf{Domain}} & \multicolumn{2}{c}{\textbf{Training}} & \multicolumn{2}{c}{\textbf{Test}} & \multirow{2}{*}{\textbf{Description}} \\
\cmidrule(lr){2-3} \cmidrule(lr){4-5}
& \textbf{Queries} & \textbf{Docs} & \textbf{Queries} & \textbf{Docs} & \\
\midrule
Biology    & 427  & 310 & 101 & 550 & Life sciences and biology textbooks and materials \\
Education  & 2,424 & 956 & 975 & 550 & Educational and pedagogical content and resources \\
Finance    & 2,490 & 689 & 315 & 550 & Financial reports, analysis, and market data \\
Government & 1,002 & 654 & 242 & 550 & Administrative, policy, and regulatory documents \\
Industrial & 2,198 & 773 & 136 & 550 & Industrial and manufacturing processes and standards \\
Legal      & 911  & 431 & 207 & 550 & Legal and regulatory compliance and contracts \\
Research   & 1,432 & 625 & 261 & 550 & Scientific and academic research publications and papers \\
\midrule
\textbf{Total} & \textbf{10,884} & \textbf{4,438} & \textbf{2,237} & \textbf{3,850} & \\
\bottomrule
\end{tabular*}
\end{table*}
\medskip

\medskip

\begingroup
\setlength{\parskip}{0.4em}
\textbf{Keyword extraction.} For each input document, we first run OCR to obtain its text content. The OCR output, together with the raw document and a rule-based prompt, is then fed into a large language model. The LLM produces a structured set of keywords for the document, capturing its main topics and terminology at the \emph{document} level rather than isolated page fragments. These keyword sets summarize document-level themes and serve as the basis for grouping documents by topical similarity in the next step.

\textbf{Document grouping.} We perform cluster analysis over the extracted keywords to form a keyword-level clustering. After sampling and document-cluster mapping, each document is assigned to one of $k$ groups, so that documents within a group are topically similar (Group~1 through Group~$k$). This grouping creates document-level semantic neighborhoods and provides the basis for generating queries that may correspond to more than one relevant document.

\textbf{Query generation.} For each group, we use an LLM with a rule prompt to generate document-level query-ground-truth (GT) pairs. The prompt guides the model to produce natural retrieval queries that are answered by one or more complete documents in the group, describing the overall topic or higher-level information needs rather than specific sentences on a single page.

\textbf{Quality validation.} We apply a two-stage validation procedure to ensure benchmark quality. In the first stage, an LLM filters candidate query-document pairs that are only weakly related, such as those linked by superficial keyword overlap or a single-page local match. In the second stage, three human annotators exhaustively review all test-split samples and manually correct problematic queries and query-document pairs to ensure document-level relevance.

For query-document relevance, the three annotators labeled 98.8\%, 99.0\%, and 94.3\% of the checked pairs as relevant, respectively, with a three-way full agreement of 92.5\% and an at-least-two agreement of 98.8\%. For query validity, the three annotators marked 99.3\%, 99.7\%, and 100.0\% of all test queries as valid. On the test set, the three-way agreement reached 99.0\%. All invalid or inconsistent samples were subsequently corrected and re-annotated before finalizing the benchmark.

Importantly, human validation serves as the final quality-control step of the benchmark, rather than a lightweight spot check. All test-split samples are exhaustively reviewed by three annotators, and any invalid, ambiguous, or weakly document-grounded cases are manually corrected before inclusion. As a result, the final benchmark is determined by human-reviewed annotations, with LLM filtering used only as a preliminary screening step.

Additional statistics on split composition and positive-target distributions are provided in Appendix~\ref{sec:appendix-dataset}. Full construction details, including LLM configurations, clustering hyperparameters, text sampling strategies, and annotation procedures, are provided in Appendix~\ref{sec:app-construction}. A query diversity analysis confirming that generated queries are not simple keyword recombinations is provided in Appendix~\ref{sec:app-query-diversity}.
\endgroup
\section{Method: DocPC}
\label{sec:method}

\subsection{Overview}

\begin{figure*}[t]
  \centering
  \includegraphics[width=\textwidth]{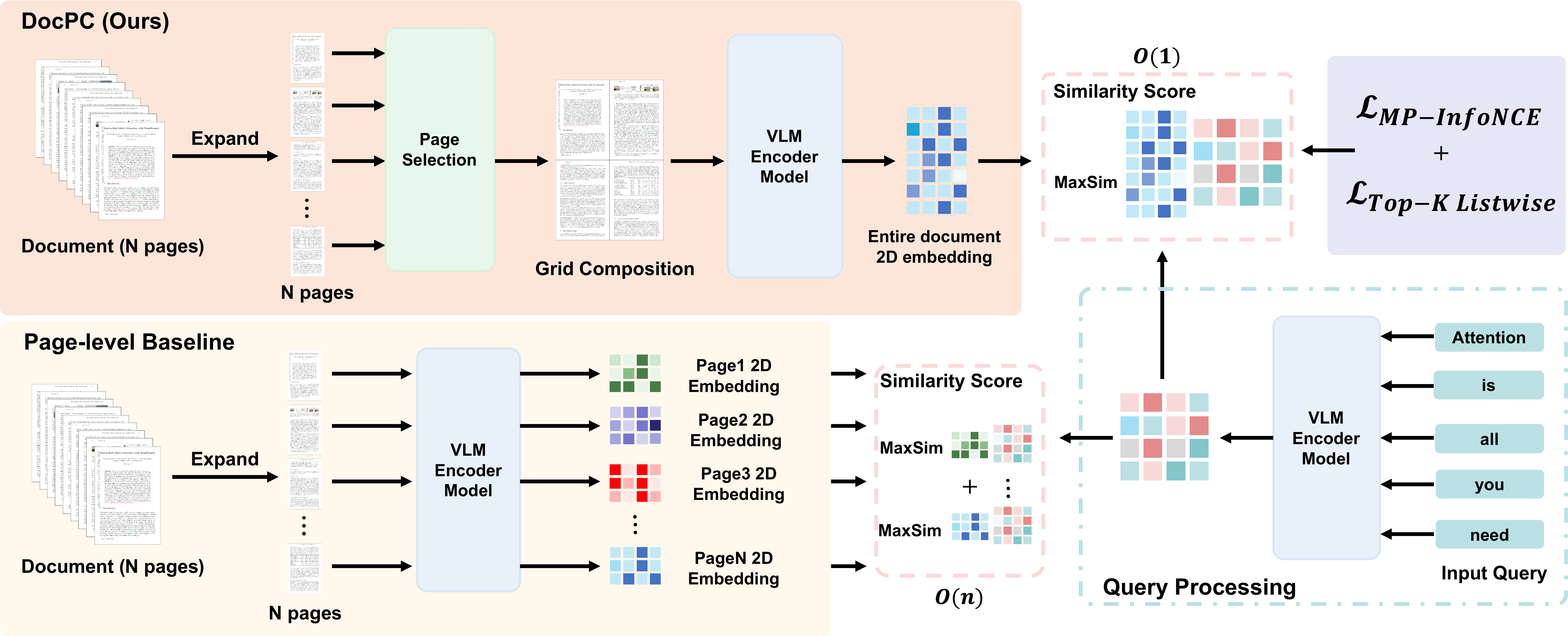}
  \caption{Overview of DocPC compared with the page-level baseline. \textbf{Top:} DocPC selects $K$ representative pages from a document, composes them into a single grid image (2$\times$2 by default), and encodes the grid once with the VLM to obtain an $\mathcal{O}(1)$ multi-vector document representation. \textbf{Bottom:} The page-level baseline encodes each page independently, producing $\mathcal{O}(N)$ embeddings per document, and aggregates per-page MaxSim scores at retrieval time. Both pipelines share the same query encoder with late interaction. Training combines Multi-Positive InfoNCE with a sparsely scheduled Top-$K$ listwise loss to handle multi-positive relevance.}
  \label{fig:main-pipeline}
\end{figure*}

DocPC represents each document by a small set of representative pages composed into a single grid image and scores it with a VLM and late interaction (Figure~\ref{fig:main-pipeline}). Given a document with $N$ pages, we select $K$ pages (Section~\ref{sec:page-composition}), compose them into a grid (2$\times$2 by default), and encode the grid once. At retrieval time we encode the query and compute similarity via late interaction (e.g., MaxSim). Training uses document-level supervision from the DocViRe training split (Section~\ref{sec:dataset}) with the objectives below. Because each document is encoded once as a fixed-size grid rather than indexed page by page, DocPC reduces document-side indexing to one representation per document. Quantitative efficiency results are reported in Section~\ref{sec:efficiency-footprint}.

\subsection{Representative Page Composition}
\label{sec:page-composition}

A core design choice is how to select $K$ pages to form the document's visual summary. Section~\ref{sec:task_formulation} shows that early pages tend to carry stronger signals. We consider both heuristic and embedding-based page selection strategies.

\textbf{First-$K$.} We take the first $K$ pages, which often contain highly informative front matter such as titles, opening sections, and tables of contents.

\textbf{Boundary.} We take the first $K/2$ and the last $K/2$ pages, targeting early and late document regions that often contain globally informative cues such as introductions, summaries, and conclusions.

\textbf{Uniform-$K$.} We sample $K$ pages at uniformly spaced indices (e.g., pages $1, \lfloor N/K \rfloor{+}1, \ldots, N$) to spread coverage across the document and avoid over-representing a single segment.

\textbf{Random-$K$.} We randomly sample $K$ pages, providing a baseline with no structural prior.

\textbf{Base-Clip.} We also consider an embedding-based dynamic page selection strategy. We first compute CLIP embeddings for all candidate pages \citep{pmlr-v139-radford21a}, cluster them, and select four representative pages from clusters to compose the final grid. Unlike fixed heuristics, this strategy selects pages according to document-specific visual-semantic structure and is used in ablation (Table~\ref{tab:page-budget}) to compare against heuristic First-4 selection.

By default we set $K=4$ and arrange the selected pages in a $2{\times}2$ grid. We use the following layouts in our experiments: $K{=}1$ gives a single-page ($1{\times}1$) representation. $K{=}4$ gives a four-page $2{\times}2$ grid. $K{=}16$ gives a 16-page larger grid. In all cases, pages are resized to a common resolution by the VLM preprocessing pipeline and arranged into a single grid image before being fed to the encoder. Thus, each document is represented by a fixed-size grid regardless of its total page count. By default we use the First-4 grid as the primary configuration. We also consider Boundary and Base-Clip as alternative page selection strategies.

\subsection{Training Objectives}
\label{sec:training-objectives}

Document-level retrieval introduces two training challenges: multiple positive documents per query (multi-positive) and the need to optimize the ranking of top-$K$ results. We address them with two losses.

\textbf{Multi-Positive InfoNCE.} Contrastive learning for visual and multimodal representations is well established \citep{pmlr-v119-chen20j,He_2020_CVPR,oord2018infonce}. Standard contrastive losses (e.g., InfoNCE) assume one positive pair per query. In document-level benchmarks like DocViRe, a query can have multiple relevant documents. Treating other positives in the batch as negatives hurts training. To relax the ``one query, one positive'' assumption, we use a multi-positive InfoNCE objective, similar in spirit to prior MP-InfoNCE formulations \citep{rivas2024conked}, that aggregates over the positive set $\mathcal{P}$ and keeps negatives in $\mathcal{N}$:
\begin{equation}
\mathcal{L}_{\mathrm{MP}\mbox{-}\mathrm{InfoNCE}} = -\log \frac{S_{\mathcal{P}}}{S_{\mathcal{P}} + S_{\mathcal{N}}},
\label{eq:mp-infonce}
\end{equation}
where $S_{\mathcal{P}} = \sum_{i \in \mathcal{P}} \exp(\mathrm{sim}(q, d_i^+) / \tau)$ and $S_{\mathcal{N}} = \sum_{j \in \mathcal{N}} \exp(\mathrm{sim}(q, d_j^-) / \tau)$. Here $\mathrm{sim}(q, d)$ is the late-interaction similarity, $\tau$ is the temperature, $d_i^+$ are positives and $d_j^-$ are negatives. This encourages the query to be close to all positives while pushed away from negatives.

\textbf{Top-$K$ listwise loss.} To directly optimize the quality of the top-$K$ ranking, we use a listwise loss based on ApproxNDCG \citep{qin2010general}, which is differentiable and focuses on the top positions:
\begin{equation}
\mathcal{L}_{\mathrm{top}\mbox{-}k} = 1 - \mathrm{ApproxNDCG}(\mathbf{y}, \mathbf{r}, K),
\label{eq:topk-loss}
\end{equation}
where $\mathbf{y}$ and $\mathbf{r}$ denote predicted scores and relevance labels.
Minimizing $\mathcal{L}_{\mathrm{top}\mbox{-}k}$ improves the ordering of the top-$K$ list relative to the relevance labels. Prior work in image retrieval has also optimized ranking metrics with listwise losses \citep{Revaud_2019_ICCV}.

\textbf{Combined objective.} We train with a weighted combination:
\begin{equation}
\mathcal{L} = \mathcal{L}_{\mathrm{MP}\mbox{-}\mathrm{InfoNCE}} + \lambda \,\mathcal{L}_{\mathrm{top}\mbox{-}k},
\label{eq:combined-loss}
\end{equation}
where $\lambda$ balances the contrastive and listwise terms. In practice, the multi-positive loss is applied at every step, while the listwise term is introduced intermittently during training. The multi-positive loss reduces false negatives among relevant documents, and the listwise loss refines the ranking of the top results.

\subsection{Training and Inference}

\textbf{Training.} We start from a pretrained page-level retrieval model ColQwen, apply the representative page composition to each document in the training set, and fine-tune on DocViRe using the combined loss $\mathcal{L}$. In-batch negatives are used for the contrastive term. No extra negative sampling is required.

\textbf{Inference.} Offline indexing is done once per document: compose a grid from the selected pages (2$\times$2 by default), run a single VLM forward pass, and store the multi-vector representation. At query time, we encode the query once, compute late-interaction scores against all document representations, and return the top-ranked documents.

\begin{table*}[!t]
  \centering
  \footnotesize
  \setlength{\tabcolsep}{4pt}
  \caption{Main results on DocViRe (NDCG@5, \%) per domain and average. Text-only baselines use OCR text or LLM-generated document summaries. Page-level VLM baselines index individual pages and aggregate with MaxP. Document-level methods use a single 4-page grid per document. For model names with two tags, the first tag denotes the training setting and the second denotes the test-time page or grid strategy. DocPC-ColQwen (First-4) uses the First-4 grid as the best configuration (see Section~\ref{sec:ablation}). Overall best in bold.}
  \label{tab:main-results}
  \begin{tabular}{@{}lrrrrrrrr@{}}
    \toprule
    Model & Bio. & Edu. & Fin. & Gov. & Ind. & Leg. & Res. & Avg \\
    \midrule
    \multicolumn{9}{@{}l}{\itshape Text-only} \\
    \midrule
    \textsc{BGE-M3} & 41.30 & 39.77 & 20.59 & 23.13 & 31.61 & 29.09 & 47.95 & 33.35 \\
    \textsc{BM25}   & 35.36 & 37.48 & 24.08 & 19.94 & 33.35 & 29.96 & 46.97 & 32.45 \\
    BGE-M3 (LLM sum.) & 36.32 & 41.07 & 15.94 & 25.92 & 34.61 & 34.19 & 50.18 & 34.03 \\
    BM25 (LLM sum.)   & 31.96 & 40.17 & 17.96 & 24.06 & 35.76 & 34.18 & 44.79 & 32.70 \\
    \midrule
    \multicolumn{9}{@{}l}{\itshape Page-level VLM baselines} \\
    \midrule
    ColQwen-page (zero-shot, All pages)& \underline{50.98} & 33.92 & 22.83 & 27.04 & 10.02 & 37.98 & 56.40 & 34.17 \\
    ColQwen-page (zero-shot, Boundary) & \textbf{52.89} & 34.20 & 22.89 & 26.54 & 8.52 & 40.58 & 56.48 & 34.59 \\
    ColQwen-page (DocViRe-sft, All pages) & 46.01 & 37.90 & 30.36 & 30.89 & 20.53 & 44.78 & \underline{56.65} & 38.16 \\
    ColQwen-page (DocViRe-sft, Boundary) & 47.09 & 38.78 & \underline{32.18} & \underline{31.93} & 20.39 & \underline{46.51} & \textbf{57.42} & 38.91 \\
    \midrule
    \multicolumn{9}{@{}l}{\itshape Document-level VLM (4-page grids)} \\
    \midrule
    Jina-CLIP (zero-shot, First-4) & 20.35 & 23.86 & 10.92 & 10.10 & 25.18 & 13.20 & 31.36 & 19.28 \\
    Nomic-Embed (zero-shot, First-4) & 4.90 & 6.96 & 8.98 & 1.74 & 16.06 & 0.73 & 10.40 & 7.11 \\
    ColQwen (DocViRe-sft, First-4) & 42.24 & \textbf{46.86} & 30.22 & 31.93 & \underline{41.11} & 42.40 & 50.47 & \underline{40.75} \\
    \midrule
    \multicolumn{9}{@{}l}{\itshape Ours} \\
    \midrule
    \textbf{DocPC-ColQwen (First-4)} & 49.02 & \underline{46.44} & \textbf{33.46} & \textbf{33.25} & \textbf{47.44} & \textbf{46.64} & 52.41 & \textbf{44.09} \\
    \bottomrule
  \end{tabular}
\end{table*}
\section{Experiments}
\label{sec:experiments}

\subsection{Setup}

We evaluate DocPC primarily on the DocViRe benchmark described in Section~\ref{sec:dataset}. DocViRe contains document-level queries and multi-positive annotations across seven domains (biology, education, finance, government, industrial, legal, and research). Unless otherwise stated, we report NDCG@5 as the main effectiveness metric and also compute Recall@$K$ on the document ranking. All metrics are computed over ranked \emph{documents} rather than pages.

\paragraph{Baselines.}
We compare DocPC against three groups of baselines that differ in retrieval unit and scoring. \emph{Text-only document retrieval} includes a sparse lexical retriever (\textsc{BM25}), a dense retriever (\textsc{BGE-M3}), and their LLM-summary variants. \emph{Page-level visual retrieval with page aggregation} includes \textsc{Jina-CLIP} \citep{koukounas2024jinaclip}, \textsc{Nomic-Embed} \citep{nussbaum2024nomicembed}, zero-shot ColQwen-page, and DocViRe-tuned ColQwen-page variants with page aggregation. \emph{Document-level visual retrieval with composed grids} includes the document-level baseline \emph{ColQwen (DocViRe-sft, First-4)} and our method \emph{DocPC-ColQwen (First-4)}. Detailed baseline construction and scoring procedures are provided in Appendix~\ref{sec:app-baselines}.

\paragraph{DocPC variants.}
DocPC represents each document by a single grid image rather than a set of individual pages. Unless otherwise stated, DocPC composes $K=4$ representative pages into a $2{\times}2$ grid and treats this grid as the retrieval unit. We use \textbf{First-4} (the first four pages) as the default strategy. In page-level contexts, ``First-4'' refers to the first four pages used in score aggregation, whereas in document-level contexts it refers to the first four pages composed into a single grid. We also consider \textbf{Boundary} (First2+Last2) as an alternative representative-page strategy. Supplementary comparisons under alternative test-time representative-page strategies are reported in Appendix~\ref{sec:app-a2b}.

\emph{Naming convention.} We distinguish two model families. \textbf{ColQwen-page (DocViRe-sft, Strategy)} is a page-level baseline fine-tuned on DocViRe with individual page images; at test time it encodes each selected page independently and aggregates scores via MaxP. \textbf{DocPC-ColQwen (Strategy)} is our document-level method fine-tuned with composed $2{\times}2$ grid images; at test time, selected pages are composed into a single grid, encoded once, and scored directly. The two families share the same VLM and late-interaction scoring but differ in input format (pages vs.\ grid), VLM passes per document ($N$ vs.\ 1), and scoring (MaxP vs.\ direct).

\paragraph{Implementation details.}
We fine-tune a publicly available ColQwen checkpoint on DocViRe using the combined objectives in Section~\ref{sec:training-objectives}, with listwise loss applied every $m{=}40$ steps (see Section~\ref{sec:ablation}). Page-level baselines use the original checkpoints in zero-shot mode unless stated otherwise; tuned ColQwen-page variants retain page-level indexing and MaxP aggregation. All methods share the same late-interaction scoring mechanism.

\subsection{Main Results}

DocPC-ColQwen (First-4) attains 44.09 average NDCG@5, versus 38.91 for the strongest page-level baseline ColQwen-page (DocViRe-sft, Boundary) under matched supervision (Table~\ref{tab:main-results}), indicating that the gain comes from document-level composition rather than more supervision. Similar gains hold under different pool sizes (Appendix~\ref{sec:app-a5}).

\subsection{Efficiency and Footprint}
\label{sec:efficiency-footprint}

Relative to page-level ColQwen on the DocViRe test set, DocPC reduces indexed images from 38{,}746 to 3{,}850 and indexed vectors from 29.25M to 2.91M (${\sim}10.1\times$). End-to-end indexing time drops from 15{,}588~s to 2{,}034~s ($7.7\times$), and Milvus storage from 22.4~GB to 2.2~GB (Appendix~\ref{sec:app-a4-efficiency}). Resolution sensitivity is analyzed in Appendix~\ref{sec:app-a4-resolution}.

\subsection{Ablation Studies}
\label{sec:ablation}

\paragraph{Grid composition vs.\ independent page encoding.}
To isolate the effect of grid composition from page selection, we compare DocPC-ColQwen (First-4 grid) with a matched independent-page baseline that uses the same four selected pages but encodes them separately and aggregates page scores with MaxP, AvgP, or SumP (Table~\ref{tab:controlled-page}).

\begin{table}[H]
  \centering
  \small
  \caption{Grid vs.\ independent page scoring (average NDCG@5, \%). All use the same First-4 pages.}
  \label{tab:controlled-page}
  \begin{tabular}{@{}llc@{}}
    \toprule
    Method & Scoring & Avg \\
    \midrule
    Independent pages (First-4) & MaxP & 37.68 \\
    Independent pages (First-4) & AvgP & 32.76 \\
    Independent pages (First-4) & SumP & 32.76 \\
    \midrule
    DocPC-ColQwen (First-4) & Grid & \textbf{44.09} \\
    \bottomrule
  \end{tabular}
\end{table}

Under the same page budget, grid composition outperforms the strongest independent-page variant (MaxP) by {+6.41} NDCG@5. MaxP selects the best single-page score, discarding complementary cross-page signals; AvgP and SumP dilute relevant pages with irrelevant ones. Grid composition instead lets the encoder jointly observe cross-page layout, headings, and topical co-occurrence in a single representation. The advantage is positive in all seven domains (Appendix~\ref{sec:app-grid-vs-independent}), with the largest gains in Industrial ({+21.08}) and Education ({+12.00}).

\paragraph{Effect of representative page budget.}
Table~\ref{tab:page-budget} (DocPC variants trained with Combined $m{=}5$, not directly comparable to Table~\ref{tab:main-results}) shows that more pages do not consistently help. A 4-page budget is the most reasonable choice in our setting.

\begin{table}[H]
  \centering
  \small
  \caption{Page budget ablation (average NDCG@5, \%). All use Combined ($m{=}5$); not directly comparable to Table~\ref{tab:main-results}.}
  \label{tab:page-budget}
  \begin{tabular}{@{}lc@{}}
    \toprule
    Model & Avg NDCG@5 \\
    \midrule
    DocPC-ColQwen (First-1) & 39.03 \\
    DocPC-ColQwen (First-4) & 40.66 \\
    DocPC-ColQwen (First-16) & 36.59 \\
    DocPC-ColQwen (Base-Clip) & 40.73 \\
    \bottomrule
  \end{tabular}
\end{table}

\paragraph{Training objectives.}
Table~\ref{tab:loss-ablation} shows the effect of training objectives and listwise loss frequency (see also Figure~\ref{fig:loss-frequency} in the Appendix). The listwise objective alone is insufficient (37.32); the multi-positive contrastive loss alone yields 42.68. The combined loss shows a three-phase pattern: at small $m$, premature ranking optimization degrades performance (40.53 at $m{=}1$); at intermediate $m$, the contrastive loss stabilizes representations before the listwise term refines ranking, peaking at 44.09 ($m{=}40$); at large $m$ (50), ranking refinement is insufficient (42.83). Additional $m$ values and a visualization are in Appendix~\ref{sec:app-loss-frequency}.

\begin{table}[H]
  \centering
  \small
  \caption{Training objective ablation (average NDCG@5, \%). Combined ($m$) applies the listwise loss every $m$ steps.}
  \label{tab:loss-ablation}
  \begin{tabular}{@{}lc@{}}
    \toprule
    Variant & Avg \\
    \midrule
    Listwise only & 37.32 \\
    Multi-positive only & 42.68 \\
    \midrule
    Combined ($m{=}1$) & 40.53 \\
    Combined ($m{=}10$) & 42.09 \\
    Combined ($m{=}40$) & \textbf{44.09} \\
    Combined ($m{=}50$) & 42.83 \\
    \bottomrule
  \end{tabular}
\end{table}

Performance by document length and a qualitative analysis are provided in Appendix~\ref{sec:app-grid-vs-independent} and~\ref{sec:qualitative}, respectively.
\section{Conclusion}

We propose DocPC, a document-level visual retriever that composes representative pages into $2{\times}2$ grids for efficient indexing, and DocViRe, a benchmark with document-grounded queries and multi-positive annotations. DocPC-ColQwen (First-4) achieves 44.09 average NDCG@5 on DocViRe, outperforming the strongest page-level baseline (38.91) while reducing indexed images, vectors, and storage by $10.1\times$. The results show that document-level composition is more effective than page-level aggregation under the same supervision.

\section{Limitations}
Grid composition inherently reduces per-page resolution; fine-grained details such as small text and dense tables can lose legibility (Section~\ref{sec:qualitative}), suggesting a hybrid grid retrieval plus page reranking approach for detail-sensitive applications. We explored only query-agnostic page selection; query-aware selection at reranking is a promising direction. DocViRe covers seven English domains; broader language and genre coverage remains for future work. The benchmark does not yet cover downstream settings such as retrieval-augmented generation.

\section{Ethics Statement}
DocViRe is constructed from the publicly available PDFA dataset \citep{pdfa_hf_2024}, which consists of openly accessible PDF documents. We manually inspected sampled documents and did not identify personally identifiable information (PII) or offensive content. Query generation was performed by a large language model (DeepSeek~V3) and subsequently validated by the paper's authors; no external crowd workers were recruited and no additional compensation was involved. The benchmark is intended solely for research on document retrieval and does not target any individual or sensitive population. We do not foresee direct negative societal impacts from our work, though we note that improved retrieval systems could in principle be applied to surveillance or privacy-invasive search if misused.

\bibliography{custom}

\clearpage
\appendix
\section{Supplementary Experimental Results}
\label{sec:appendix-results}

In this appendix we use the same terminology as in the main paper: \emph{page-level} retrieval indexes each page separately and aggregates scores at query time. \emph{document-level} retrieval composes a grid per document and encodes it once. We distinguish \textbf{First-4 pages} (first four pages used in page-level aggregation) from \textbf{First-4 grid} (first four pages composed into a single document-level grid). All NDCG@5 values are reported in percentage (0 to 100).

\subsection{Cross-Strategy Generalization of DocPC}
\label{sec:app-a2}

Table~\ref{tab:app-a3} reports cross-strategy generalization of DocPC-ColQwen. Rows denote training page-selection strategies and columns denote test grid strategies. Values are average NDCG@5 (\%) over the seven domains.

\begin{table}[htbp]
  \centering
  \footnotesize
  \setlength{\tabcolsep}{4pt}
  \caption{Cross-strategy generalization of DocPC-ColQwen (average NDCG@5, \%). Rows denote training strategies and columns denote test strategies. F4 = First-4, Bnd. = Boundary, U4 = Uniform-4, R4 = Random-4, and L4 = Last-4.}
  \label{tab:app-a3}
  \begin{tabular}{@{}lccccc@{}}
    \toprule
    Train $\backslash$ Test & F4 & Bnd. & U4 & R4 & L4 \\
    \midrule
    Bnd. & 42.78 & 40.75 & 40.34 & 38.62 & 36.33 \\
    F4   & 40.66 & 39.17 & 38.95 & 36.77 & 35.36 \\
    L4   & 40.23 & 38.28 & 38.72 & 37.00 & 34.93 \\
    R4   & 42.64 & 40.69 & 40.17 & 39.60 & 36.00 \\
    U4   & \textbf{43.70} & 41.33 & 41.29 & 38.56 & 36.90 \\
    \bottomrule
  \end{tabular}
\end{table}

\subsection{Comparison of Tuned ColQwen and DocPC Under Different Test-Time Strategies}
\label{sec:app-a2b}

Table~\ref{tab:app-a3b} compares tuned ColQwen and DocPC-ColQwen under different test-time representative-page strategies. All values are average NDCG@5 (\%) over the seven domains.

\begin{table}[t]
  \centering
  \footnotesize
  \setlength{\tabcolsep}{4pt}
  \caption{Tuned ColQwen vs.\ DocPC-ColQwen under different test-time representative-page strategies on DocViRe (average NDCG@5, \%).}
  \label{tab:app-a3b}
  \begin{tabular}{@{}llc@{}}
    \toprule
    Method & Strategy & Avg \\
    \midrule
    ColQwen (DocViRe-sft) & First-4   & 40.75 \\
    ColQwen (DocViRe-sft) & Boundary  & 39.26 \\
    ColQwen (DocViRe-sft) & Uniform-4 & 37.96 \\
    ColQwen (DocViRe-sft) & Random-4  & 36.23 \\
    \midrule
    DocPC-ColQwen         & First-4   & \textbf{43.70} \\
    DocPC-ColQwen         & Boundary  & 41.33 \\
    DocPC-ColQwen         & Uniform-4 & 41.29 \\
    DocPC-ColQwen         & Random-4  & 39.60 \\
    \bottomrule
  \end{tabular}
\end{table}

\subsection{Extended Ablations on Training Objective and Page Budget}
\label{sec:app-a4}

The following tables are \emph{extended/exploratory} ablations. Configurations and training setup may differ from the main paper. Values are not directly comparable to Table~\ref{tab:main-results} or the page-budget table in Section~\ref{sec:ablation}, which use controlled settings. The results suggest that the benefit of the listwise term depends strongly on how frequently it is applied. A possible explanation is that the multi-positive contrastive loss primarily shapes the representation space, whereas the listwise ApproxNDCG term mainly refines the ordering among already competitive candidates. When applied too frequently, the listwise term may interfere with early-stage representation learning; when applied sparsely, it can serve as a lightweight ranking refinement on top of a stronger representation space.

\begin{table}[htbp]
  \centering
  \footnotesize
  \caption{Extended loss-frequency ablation for DocPC-ColQwen (NDCG@5, \% AVG). Exploratory. Not directly comparable to main-text Table~\ref{tab:loss-ablation} (controlled configuration). Listwise only / Multi-positive only and Combined($m$) with various $m$.}
  \label{tab:app-a6}
  \begin{tabular}{@{}lc@{}}
    \toprule
    Variant & Avg \\
    \midrule
    Combined (m=1) & 40.53 \\
    Combined (m=3) & 41.61 \\
    Combined (m=5) & 40.66 \\
    Combined (m=7) & 42.70 \\
    Combined (m=9) & 42.96 \\
    Combined (m=11) & 43.18 \\
    Combined (m=13) & 43.67 \\
    Combined (m=15) & 43.83 \\
    Combined (m=20) & 43.36 \\
    Combined (m=30) & 43.27 \\
    Combined (m=40) & \textbf{44.09} \\
    Combined (m=50) & 42.83 \\
    Listwise only & 37.32 \\
    Multi-positive only & 42.68 \\
    \bottomrule
  \end{tabular}
\end{table}

\begin{table}[htbp]
  \centering
  \footnotesize
  \caption{Extended representative-page budget ablation for DocPC-ColQwen (NDCG@5, \% AVG). Exploratory. Main-text page-budget table (Section~\ref{sec:ablation}) uses a different controlled setup (First-1/First-4/First-16, CLIP-selected 4-page grid).}
  \label{tab:app-a7}
  \begin{tabular}{@{}lc@{}}
    \toprule
    Model & Avg \\
    \midrule
    DocPC-ColQwen (First-1) & 39.03 \\
    DocPC-ColQwen (First-4) & 40.66 \\
    DocPC-ColQwen (First-16) & 36.59 \\
    DocPC-ColQwen (CLIP-selected 4-page grid) & 40.73 \\
    \bottomrule
  \end{tabular}
\end{table}


\subsection{Detailed efficiency breakdown}
\label{sec:app-a4-efficiency}

Table~\ref{tab:app-efficiency-footprint} reports the detailed efficiency and storage comparison between DocPC and page-level ColQwen on the DocViRe test index.
DocPC reduces the number of indexed images and vectors by about $10.1\times$, lowers Milvus storage from 22.4~GB to 2.2~GB, and reduces end-to-end indexing time from 15{,}588~s to 2{,}034~s.
While the composed 4-page grid slightly increases per-image encoding time, the overall encoding and insertion costs are substantially reduced because far fewer inputs need to be indexed.

\begin{table}[htbp]
  \centering
  \footnotesize
  \setlength{\tabcolsep}{4pt}
  \caption{Detailed efficiency and storage comparison between DocPC and page-level ColQwen on the DocViRe test index.}
  \label{tab:app-efficiency-footprint}
  \begin{tabular}{@{}lccc@{}}
    \toprule
    Metric & DocPC & ColQwen & Ratio \\
    \midrule
    Images   & 3{,}850   & 38{,}746 & $10.1\times$ fewer \\
    Vectors  & 2.91M     & 29.25M   & $10.1\times$ fewer \\
    Enc./img & 431.7 ms  & 304.7 ms & 1.42$\times$ higher \\
    Enc.     & 1{,}663 s & 11{,}805 s & $7.1\times$ faster \\
    Insert   & 371 s     & 3{,}783 s & $10.2\times$ faster \\
    Total    & 2{,}034 s & 15{,}588 s & $7.7\times$ faster \\
    Storage  & 2.2 GB    & 22.4 GB  & $10.1\times$ smaller \\
    \bottomrule
  \end{tabular}
\end{table}

\subsection{Resolution sensitivity}
\label{sec:app-a4-resolution}

We further examine how DocPC-ColQwen (First-4 grid) behaves under different input resolutions of the composed grid image on DocViRe. As shown in Table~\ref{tab:app-a7b}, average NDCG@5 rises from 6.68 at $128{\times}166$ to 22.13 at $256{\times}331$, 37.55 at $512{\times}662$, and 40.65 at the original-resolution setting, indicating consistent but diminishing gains as resolution increases.

\begin{table}[htbp]
  \centering
  \footnotesize
  \caption{Resolution sensitivity of DocPC-ColQwen (First-4 grid) on DocViRe (average NDCG@5, \%).}
  \label{tab:app-a7b}
  \begin{tabular}{@{}lc@{}}
    \toprule
    Resolution & Avg NDCG@5 \\
    \midrule
    $128{\times}166$ & 6.68 \\
    $256{\times}331$ & 22.13 \\
    $512{\times}662$ & 37.55 \\
    Original         & 40.65 \\
    \bottomrule
  \end{tabular}
\end{table}

\subsection{Retrieval Performance Under Varying Candidate-Pool Sizes}
\label{sec:app-a5}

Table~\ref{tab:app-a8} reports retrieval performance as the candidate document pool size varies (per-domain pool size: 550, 400, or 200 size). The page-level baseline is ColQwen-page with MaxP aggregation (taking the better of the two ColQwen-page variants for each pool size). DocPC-ColQwen (First-4) uses the First-4 grid. ``Gain'' is the absolute difference in average NDCG@5 (percentage points) between DocPC and the page-level baseline.

\begin{table}[htbp]
  \centering
  \footnotesize
  \caption{Retrieval performance under varying candidate-pool sizes (NDCG@5, \% AVG). Page-level baseline: ColQwen-page. Gain = DocPC minus page-level (absolute percentage points).}
  \label{tab:app-a8}
  \begin{tabular}{@{}lccc@{}}
    \toprule
    Pool size & Page-level & DocPC-ColQwen (First-4) & Gain \\
    \midrule
    550 & 40.75 & 44.09 & 3.34 \\
    400 & 41.19 & 44.28 & 3.09 \\
    200 & 50.15 & 52.29 & 2.14 \\
    \bottomrule
  \end{tabular}
\end{table}

\subsection{Baseline Construction and Scoring Details}
\label{sec:app-baselines}

For text-only baselines, we first extract OCR text from each document. \textsc{BM25} is applied directly to the extracted text, while \textsc{BGE-M3} encodes OCR-derived text chunks into dense embeddings. We also include LLM-summary variants where each document is first compressed into a single abstractive summary and retrieval is performed over summaries using \textsc{BGE-M3} or \textsc{BM25}. OCR outputs are split into fixed-size chunks, each chunk is encoded once, and document scores are obtained by taking the maximum similarity over all chunks (MaxP aggregation).

For page-level visual baselines, each page is treated as a retrieval unit. We compute query-page similarities with late interaction and obtain document scores via MaxP over pages, ensuring a uniform aggregation scheme across page-level methods. We report zero-shot ColQwen-page and ColQwen-page (Boundary), as well as DocViRe-tuned page-level variants: \emph{ColQwen-page (DocViRe-sft, All pages)} and \emph{ColQwen-page (DocViRe-sft, Boundary)}, which use the same supervision as DocPC while retaining page-level indexing and MaxP aggregation during inference.

For document-level bi-encoder baselines, \textsc{Jina-CLIP} and \textsc{Nomic-Embed} are evaluated with First-4 grids, so that each document is represented by a single composed image.

\section{Supplementary Dataset Statistics}
\label{sec:appendix-dataset}

\subsection{Dataset Collection and Preprocessing}
To construct DocViRe, we used the first 40 source files to build the training split and files 41 to 50 to build the test split. After obtaining the raw PDFs, we first retained only documents with 4 to 30 pages. We then applied DeepSeek-based document classification and kept only those PDFs assigned to one of the seven target domains: biology, education, finance, government, industrial, legal, and research. The same preprocessing and filtering procedure was applied independently to the test split. This classification step was used only for corpus filtering, rather than as benchmark relevance annotation.

\subsection{Quality validation agreement}
For the test split, the three annotators labeled 98.8\%, 99.0\%, and 94.3\% of checked query-document pairs as relevant, with a three-way full agreement of 92.5\% and an at-least-two agreement of 98.8\%. For query validity, they exhaustively reviewed all test-split queries and marked 99.3\%, 99.7\%, and 100.0\% of them as valid. Across all test queries that received validity labels from all three annotators, the three-way agreement reached 99.0\%.

\subsection{Construction Details}
\label{sec:app-construction}

We provide the complete specification of the DocViRe construction pipeline.

\textbf{LLM and API details.} All LLM-based steps use DeepSeek~V3 (deepseek-v3-2-251201) accessed via the Volcengine API. Specific configurations per step:
\begin{itemize}
  \item \emph{Keyword extraction:} temperature=0.3, max\_tokens=1024. Each document is sampled to 8{,}000 characters (first 2{,}000 + 8 evenly spaced 500-character blocks from the middle + last 2{,}000) before being sent to the model.
  \item \emph{Query generation:} temperature=0.3, max\_tokens=500. The prompt receives the shared cluster keywords and partial excerpts (same 8{,}000-character sampling) from up to 5 documents in the group. The prompt explicitly instructs: ``Make it sound like a typical search. Avoid technical phrasing, abstract concepts, or long constraints.''
  \item \emph{LLM reverse validation:} temperature=0.2, max\_tokens=2{,}000. The full document text (up to 150{,}000 characters) is provided along with all candidate queries. The model judges which queries are semantically relevant. Queries linked only by superficial keyword overlap are filtered out.
\end{itemize}

\textbf{Clustering details.} Keywords are embedded using Qwen3-Embedding-8B. We apply FAISS KMeans clustering with n\_iter=25 and seed=42. The number of clusters $k$ is set to $\sqrt{n}$ (number of unique keywords), clamped to $[2, 100]$. We also use the elbow method (SSE vs.\ $k$) as a sanity check, though $\sqrt{n}$ generally provides a reasonable default. After clustering, each document is assigned the cluster IDs of its keywords. Documents sharing memberships in multiple clusters (cluster\_num ranging from 2 to 10) are grouped together, with each group containing 3--5 documents.

\textbf{Annotation procedure.} All annotations were performed by three of the paper's authors. No external annotators were recruited, and no additional compensation was involved. For the test split, all query-document pairs and query validity labels were exhaustively reviewed by all three annotators. Specifically, for each test query, annotators reviewed the entire candidate pool (all documents in the test split), not just the LLM-generated positives. Invalid, ambiguous, or weakly grounded cases were corrected through discussion and re-annotation. The three-way full agreement reached 92.5\% for relevance and 99.0\% for query validity (Section~\ref{sec:dataset}).

\subsection{Query Diversity Analysis}
\label{sec:app-query-diversity}

To verify that the generated queries are not simple keyword recombinations, we measure the unigram lexical overlap between each test query and the document-level keywords of its relevant documents (Table~\ref{tab:query-overlap}).

\begin{table}[htbp]
  \centering
  \footnotesize
  \caption{Query-keyword lexical overlap statistics on the DocViRe test set.}
  \label{tab:query-overlap}
  \begin{tabular}{@{}p{0.7\columnwidth}r@{}}
    \toprule
    Metric & Value \\
    \midrule
    Avg.\ Jaccard similarity & 0.07 \\
    Avg.\ query coverage & 41.6\% \\
    Avg.\ novel ratio & 58.4\% \\
    Queries w/ zero overlap & 5.9\% \\
    Queries w/ coverage ${\leq}50\%$ & 71.7\% \\
    Queries w/ 100\% coverage & 1.8\% \\
    \bottomrule
  \end{tabular}
\end{table}

The average Jaccard similarity is only 0.07, confirming that queries and document keywords occupy largely disjoint lexical spaces. On average, 58.4\% of query unigrams do not appear in any associated document keyword, and 71.7\% of queries have ${\leq}50\%$ token coverage. This confirms that the three-stage generation pipeline (group-level LLM generation $\rightarrow$ reverse LLM validation $\rightarrow$ exhaustive human review) produces queries with natural lexical diversity rather than keyword-copying bias.

\subsection{Per-Domain Grid vs.\ Independent Page Comparison}
\label{sec:app-grid-vs-independent}

Table~\ref{tab:app-grid-domain} reports the per-domain NDCG@5 comparison between DocPC grid and independent pages with MaxP under the matched First-4 setting. Table~\ref{tab:app-grid-length} reports the comparison by document length.

\begin{table}[htbp]
  \centering
  \footnotesize
  \caption{Per-domain NDCG@5 (\%) comparison between DocPC grid and independent pages with MaxP.}
  \label{tab:app-grid-domain}
  \begin{tabular}{@{}lrrr@{}}
    \toprule
    Domain & DocPC grid & Indep.\ MaxP & $\Delta$ \\
    \midrule
    Biology    & 46.62 & 46.47 & +0.15 \\
    Education  & 50.37 & 38.37 & +12.00 \\
    Finance    & 33.52 & 28.60 & +4.92 \\
    Government & 31.72 & 26.89 & +4.83 \\
    Industrial & 40.98 & 19.90 & +21.08 \\
    Legal      & 47.66 & 46.66 & +1.00 \\
    Research   & 58.03 & 56.85 & +1.18 \\
    \bottomrule
  \end{tabular}
\end{table}

The grid advantage is positive in all seven domains. The gains are largest in visually rich domains such as Industrial ({+21.08}) and Education ({+12.00}), where document-level evidence is often distributed across headers, tables, figures, and short text blocks on multiple early pages. In text-heavy domains with more uniform layouts (Biology, Legal, Research), both methods perform comparably, suggesting that independent high-resolution page encoding can preserve fine-grained textual details while grid composition is most helpful when cross-page visual context matters.

\begin{table}[htbp]
  \centering
  \footnotesize
  \caption{DocPC grid vs.\ independent pages (MaxP) by document length (NDCG@5, \%).}
  \label{tab:app-grid-length}
  \begin{tabular}{@{}lrrrr@{}}
    \toprule
    Document Length & \#Queries & Grid & Page & $\Delta$ \\
    \midrule
    $\leq$10 pages & 1{,}331 & 43.82 & 36.75 & +7.07 \\
    11--20 pages   & 765     & 48.05 & 38.84 & +9.21 \\
    $>$20 pages    & 141     & 53.58 & 44.01 & +9.57 \\
    \bottomrule
  \end{tabular}
\end{table}

The grid advantage grows from {+7.07} for short documents to {+9.57} for long documents. Longer documents contain more diverse content across pages, and the grid's ability to jointly encode multiple pages becomes increasingly valuable.

Table~\ref{tab:length-analysis} reports DocPC-ColQwen (First-4) performance grouped by relevant-document page count. Performance is highest for mid-length documents (11--20 pages), while short documents (${\leq}10$ pages) yield the lowest NDCG@5. This reflects pool composition: short documents account for 59.5\% of queries, increasing topical overlap and retrieval difficulty.

\begin{table}[htbp]
  \centering
  \footnotesize
  \caption{DocPC performance by document length (NDCG@5, \%).}
  \label{tab:length-analysis}
  \begin{tabular}{@{}lrrr@{}}
    \toprule
    Group & Docs & Queries & NDCG@5 \\
    \midrule
    $\leq$10 pages  & 570 & 1721 & 37.57 \\
    11 to 20 pages  & 248 & 988 & 44.55 \\
    $>$20 pages     & 79 & 317 & 40.62 \\
    \bottomrule
  \end{tabular}
\end{table}

\subsection{Page Selection Strategy Analysis}
\label{sec:app-page-strategy}

We analyze the interaction between page selection strategy and document characteristics. Table~\ref{tab:app-info-dist} classifies queries by their information distribution pattern. For each query, we compare the NDCG@5 achieved by First-4, Last-4, and Uniform-4. If First-4 substantially outperforms Last-4 ($\Delta > 0.2$), the document's relevant content is likely concentrated in the early pages (``front-heavy''); the reverse indicates ``end-heavy''; if Uniform-4 performs best, the information is distributed throughout (``balanced'').

\begin{table}[htbp]
  \centering
  \scriptsize
  \setlength{\tabcolsep}{3pt}
  \caption{NDCG@5 ($\times$100) by information distribution pattern.}
  \label{tab:app-info-dist}
  \begin{tabular}{@{}lrccccc@{}}
    \toprule
    Pattern & \#Q & F4 & Bnd. & U4 & L4 \\
    \midrule
    Front-heavy & 582  & \textbf{68.17} & 55.90 & 59.08 & 19.60 \\
    Balanced    & 1{,}108 & 39.09 & 41.69 & \textbf{46.53} & 38.67 \\
    End-heavy   & 344  & 22.29 & 47.54 & 39.61 & \textbf{64.29} \\
    \bottomrule
  \end{tabular}
\end{table}

First-4 is the single best strategy for 39.7\% of individual queries and benefits from the empirical regularity that key document information (titles, abstracts, tables of contents) tends to appear early. For corpora known to contain long, content-distributed documents (e.g., textbooks, technical manuals), Uniform-4 provides better coverage. Boundary is effective for report-style documents with executive summaries and conclusions. For deployment in domains with known structural conventions, practitioners can select strategies accordingly.

\subsection{Loss Frequency Visualization}
\label{sec:app-loss-frequency}

\begin{figure}[htbp]
  \centering
  \includegraphics[width=\columnwidth]{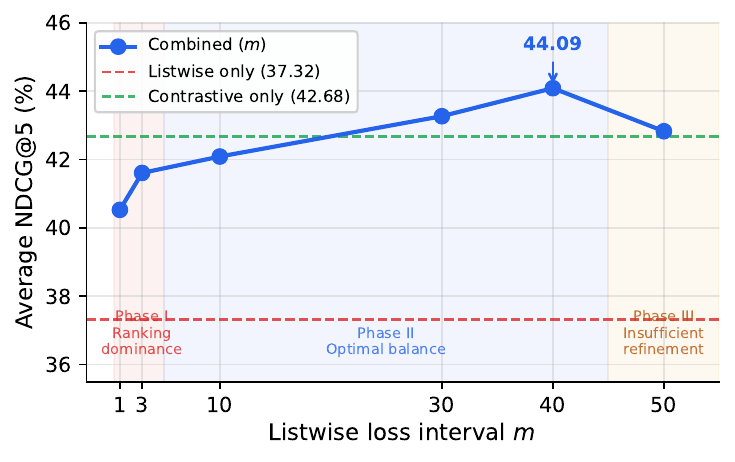}
  \caption{Effect of listwise loss interval $m$ on NDCG@5. The curve reveals a three-phase pattern: ranking loss dominance ($m{\leq}3$), optimal balance ($m{=}10$--$40$), and insufficient ranking refinement ($m{=}50$). Dashed lines show standalone baselines.}
  \label{fig:loss-frequency}
\end{figure}

\subsection{Qualitative Analysis}
\label{sec:qualitative}

To understand the strengths and limitations of grid composition, we compare DocPC-ColQwen (First-4, Grid) with its page-level counterpart (Independent, First-4, MaxP) across all 2{,}237 test queries (Table~\ref{tab:outcome}).

\begin{table}[htbp]
  \centering
  \footnotesize
  \caption{Outcome distribution across 2{,}237 test queries. A method ``succeeds'' if it retrieves at least one relevant document in the top 5.}
  \label{tab:outcome}
  \begin{tabular}{@{}lrr@{}}
    \toprule
    Outcome & Count & \% \\
    \midrule
    Both succeed & 1{,}108 & 49.5 \\
    Grid only succeeds & 398 & 17.8 \\
    Page only succeeds & 218 & 9.7 \\
    Both fail & 513 & 22.9 \\
    \bottomrule
  \end{tabular}
\end{table}

Grid composition exclusively succeeds in 17.8\% of queries while exclusively failing in 9.7\%, yielding a 1.83$\times$ net positive ratio. The grid advantage is most pronounced for queries describing cross-page themes (e.g., ``student complaints procedure university''), where the complaint procedure spans multiple early pages with headers, policy text, and flowcharts. The $2{\times}2$ grid preserves the co-occurrence of these elements in a single representation, enabling the model to match the query holistically, while no single page contains sufficient signal for independent encoding.

Conversely, grid compression loses critical information when documents contain fine-grained visual details. For queries targeting medical forms with dosage tables or dense patient intake forms, the small-font text becomes illegible in the grid. In these cases, independent page encoding preserves per-page resolution and retrieves the correct document.

Queries that defeat both methods typically involve very long documents (${\geq}30$ pages) where the relevant content lies beyond the first 4 pages, reflecting a fundamental limitation of the First-4 page selection strategy rather than the grid representation itself.

We acknowledge that grid composition is not uniformly beneficial. The 218 page-only success cases (9.7\%) represent genuine information loss due to resolution reduction. However, this is outweighed by the 398 grid-only successes (17.8\%), where joint encoding captures cross-page semantics that independent encoding misses. For applications requiring fine-grained visual detail, a hybrid approach combining grid-level retrieval with page-level reranking is a promising direction.

\end{document}